\documentclass[11pt,twocolumn,twoside,a4paper,amsmath,amssymb,aps,showkeys,showpacs]{revtex4}

\usepackage{amsfonts}
\usepackage{graphics} 
\usepackage{epsfig}
\usepackage{fancyheadings}

\def\bea{\begin{eqnarray}}
\def\eea{\end{eqnarray}}

\begin{document}
\thispagestyle{myheadings}
\rhead[]{}
\lhead[]{}
\chead[Thomas A. Trainor]{Fragmentation systematics in nuclear collisions}

\title{Systematics of parton fragmentation  in $e^+$-$e^-$ and  nuclear collisions}

\author{Thomas A. Trainor}

\affiliation{%
CENPA 354290, University of Washington, Seattle, WA
 98195,  USA 
}%

\received{?}

\begin{abstract}
Parametrizations of fragmentation functions (FFs) from $e^+$-$e^-$ and p-\=p collisions are combined with a parton spectrum model in a pQCD folding integral to produce minimum-bias {\em fragment distributions}. A model of in-medium FF modification is included.  Calculated fragment distributions are compared with {\em hard components} from p-p and Au-Au $p_t$ spectra. Data are well described by pQCD over a large kinematic region for a range of Au-Au centralities.
\end{abstract}



\maketitle

\renewcommand{\thefootnote}{\fnsymbol{footnote}}
\renewcommand{\thefootnote}{\roman{footnote}}


\section{Introduction}     \label{intro}

RHIC collisions are conventionally  described in terms of hydrodynamic (hydro) evolution of a thermalized bulk medium and energy loss of energetic partons (hard probes) in that medium. Hydro should dominate $p_t$ spectra below 2 GeV/c, parton fragmentation above 5 GeV/c, and ``quark coalescence'' in the intermediate $p_t$ interval.

However, recent analysis of spectrum and correlation structure has revealed {\em minijet} structures 
in RHIC collisions~\cite{ppcorr1,ppcorr2,axialci,hijscale,ptscale,ptedep,daugherity}. Two-component analysis of \mbox{p-p} and Au-Au spectra reveals a corresponding {\em hard component} (minimum-bias {\em fragment distribution}), suggesting that jet phenomena extend down to 0.1 GeV/c~\cite{ppprd,hardspec}. Minijets~\cite{minijet} appear to dominate the transverse dynamics of nuclear collisions at energies above $\sqrt{s_{NN}} \sim$ 15 GeV and provide {unbiased} access to fragment distribution structure down to a small cutoff energy for scattered partons (3 GeV)  and to the smallest detectable fragment momenta ($\sim 0.1$ GeV/c). 

Minijets can be studied in the form of $p_t$-spectrum hard components isolated via the two-component spectrum model. Measured hard components are compared with pQCD fragment distributions (FDs). Parton spectrum parameters and modifications to fragmentation functions (FFs) in more-central Au-Au collisions are inferred~\cite{evolve}. The goal is a comprehensive pQCD description of all nuclear collisions.

\section{Two-component  model}

The two-component (soft+hard) spectrum model was first obtained from a Taylor-series expansion of p-p $p_t$ spectra on {uncorrected} event multiplicity $\hat n_{ch}$ for ten multiplicity classes~\cite{ppprd}. The soft component was interpreted as longitudinal nucleon fragmentation, the hard component as transverse scattered-parton fragmentation.


The two-component model for p-p collisions with soft and hard multiplicities $n_s + n_h = n_{ch}$ is
\bea \label{twoc}
 \frac{1}{n_s(\hat n_{ch})}\frac{1}{y_t}\, \frac{dn_{ch}(\hat n_{ch})}{dy_t } =  S_0(y_t)  +  \frac{n_h(\hat n_{ch})}{n_s(\hat n_{ch})}\,  H_{0}(y_t),
\eea
Coefficient $n_h  / n_s$ scales as $\alpha\, \hat n_{ch}$,
$S_0(y_t)$ is a L\'evy distribution on $m_t$ and $H_0(y_t)$ is a Gaussian plus QCD power-law tail on transverse rapidity $y_t = \ln\{(m_t + p_t) / m_0\}$~\cite{ppprd}. To compare with A-A spectra we define $S_{pp} =(1/y_t)\, dn_s/dy_t$ with reference model $n_s\, S_0$ and similarly for $H_{pp} \leftrightarrow n_h\, H_0$. 


\begin{figure*}[th]
\includegraphics[height=1.65in]{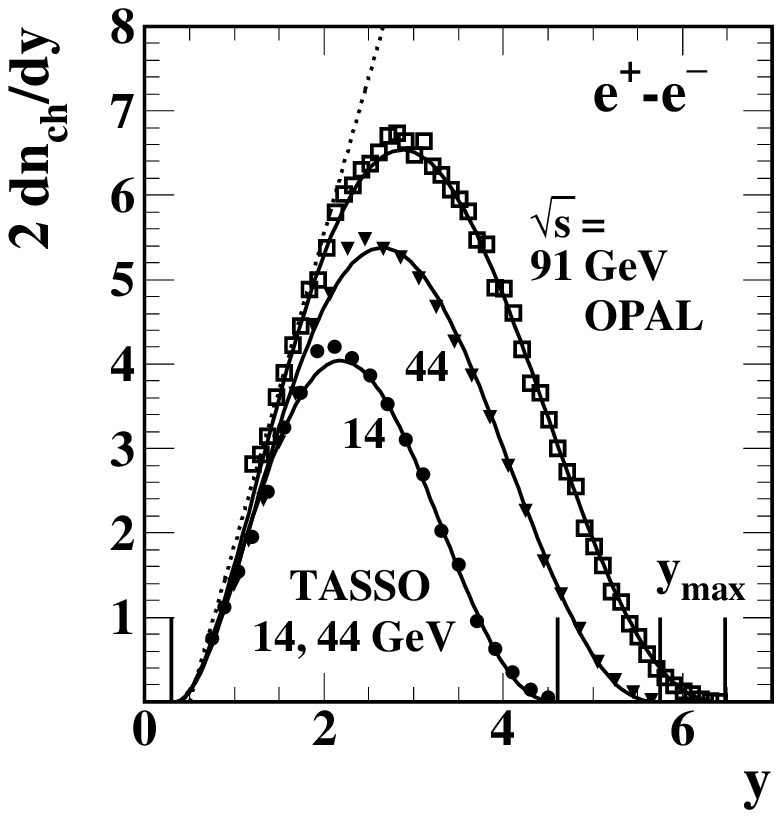} 
\includegraphics[height=1.65in]{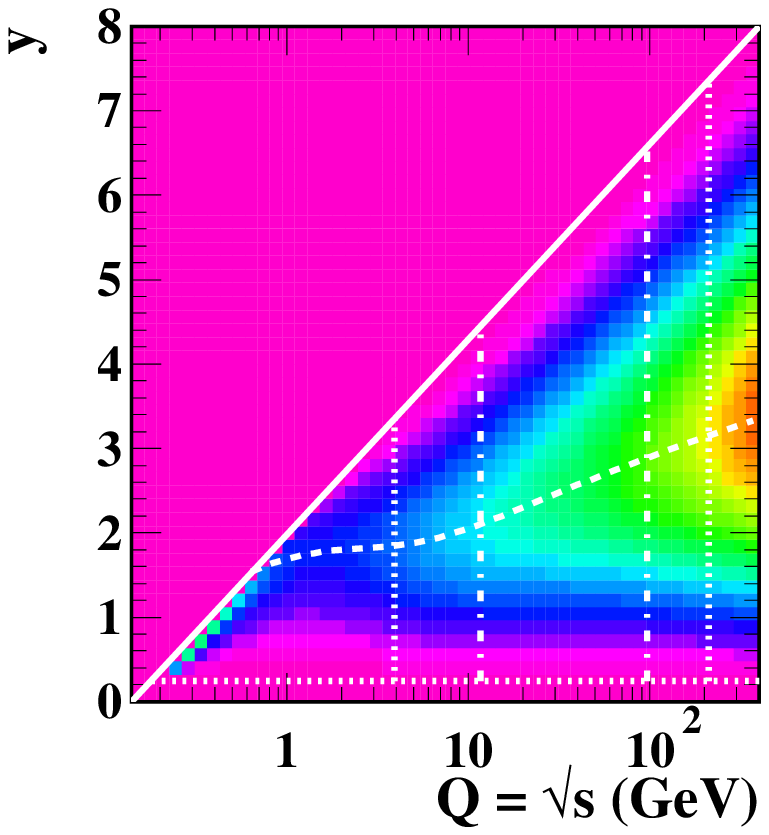}
\includegraphics[height=1.67in]{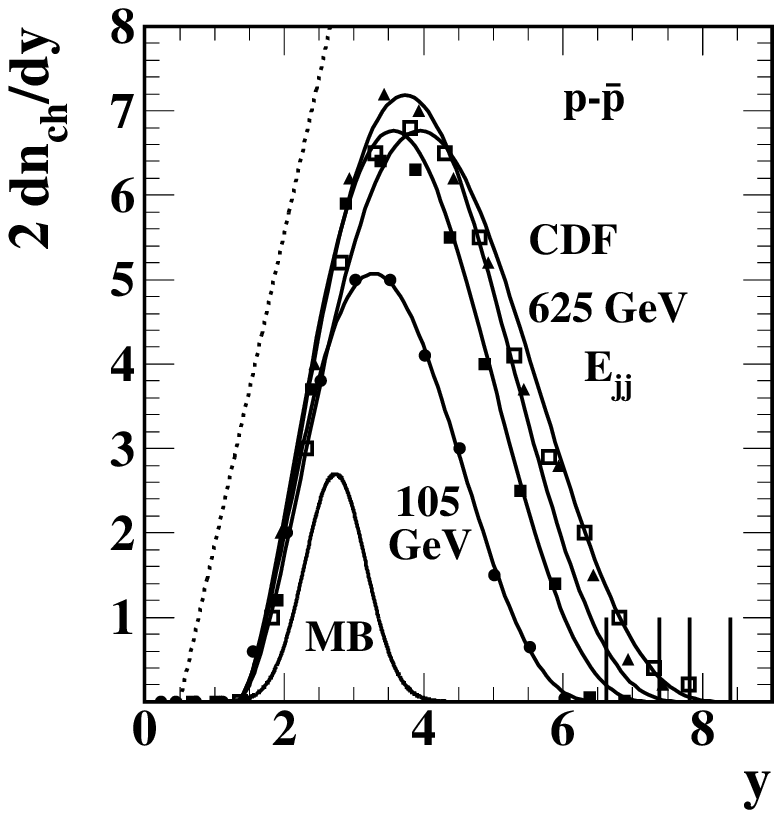}
\includegraphics[height=1.65in]{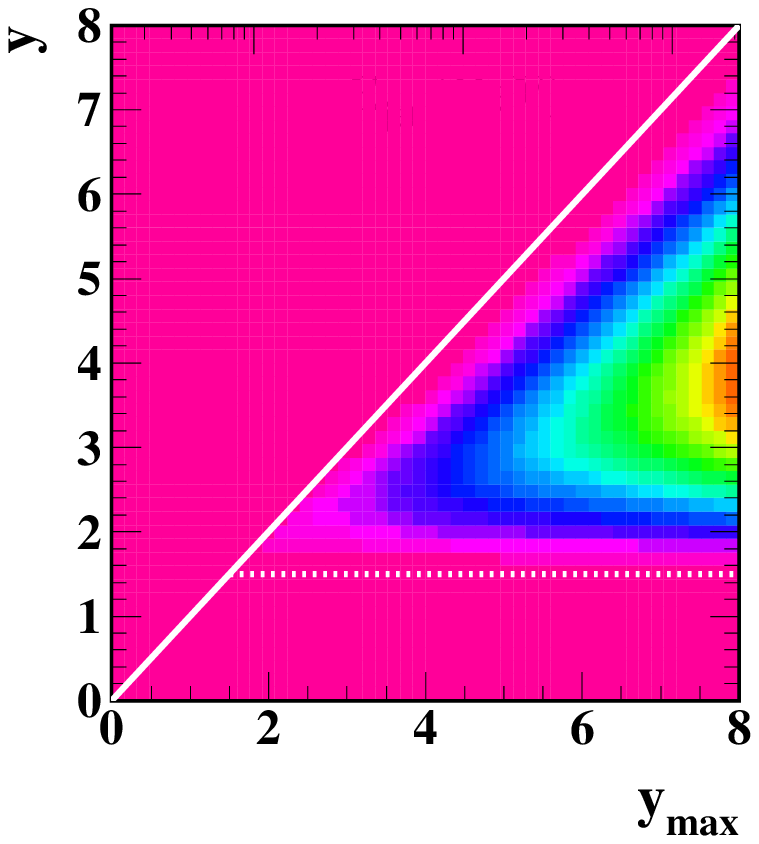}
\caption{\label{ttfig1}
First: Fragmentation functions (FFs) from $e^+$-$e^-$ collisions for three energies with $\beta$-distribution parametrizations (solid curves),
Second: Full $e^+$-$e^-$ FF parametrization on parton rapidity $y_{max}$,
Third: FFs from p-\=p collisions for several dijet energies,
Fourth: Full p-\=p FF parameterization on parton rapidity.
}
\end{figure*}


The corresponding two-component model for per-participant-pair A-A spectra is 
\bea  \label{aa2comp}
\frac{2}{n_{part}} \frac{1}{y_t}\frac{dn_{ch}}{dy_t} &=& S_{NN}(y_t) +  \nu\, H_{AA}(y_t;\nu) \\ \nonumber
&=&  S_{NN}(y_t) +  \nu\,r_{AA}(y_t;\nu) \,H_{NN}(y_t),
\eea
where $S_{NN}$ ($\sim S_{pp}$) is the soft component and $H_{AA}$ is the A-A hard component (with reference $H_{NN}\sim H_{pp}$) 
~\cite{ppprd,hardspec}. Ratio $r_{AA} = H_{AA} / H_{NN}$ is an alternative to nuclear modification factor $R_{AA}$. Centrality measure $\nu \equiv 2 n_{binary} / n_{participant}$ estimates the 
mean nucleon path length. We are interested in the evolution of hard component $H_{AA}$ or ratio $r_{AA}$ with A-A centrality.

\section{Fragmentation functions}

$e^+$-$e^-$ (e-e) fragmentation functions (FFs) have been parametrized accurately over the full kinematic region relevant to nuclear collisions~\cite{ffprd}. Light-quark and gluon fragmentation functions\, $ D_{xx}(x,Q^2) \leftrightarrow D_{xx}(y,y_{max}) $ ($xx$ = \mbox{e-e}, p-p, A-A) are described above energy scale $Q =2\, E_{jet} \sim 10$ GeV by a two-parameter {\em beta distribution} $\beta(u;p,q)$ on normalized rapidity $u$~\cite{ffprd}. Fragment rapidity for unidentified hadrons is $y = \ln[(E+p)/m_\pi]$, and parton rapidity $y_{max} = \ln(Q/m_\pi)$. Parameters $(p,q)$ vary slowly and linearly with $y_{max}$ above $Q = 10$ GeV and can be extrapolated down to $Q \sim 4$ GeV. 

Fig.~\ref{ttfig1} (first panel) shows measured FFs for three energy scales from HERA/LEP~\cite{tasso,opal}. The curves are $\beta(p,q)$ parametrizations which describe data over the entire fragment momentum range. 
Fig.~\ref{ttfig1} (second panel) shows the FF ensemble 
vs energy scale $Q$ as a surface plot~\cite{ffprd}. 

Figure~\ref{ttfig1} (third panel) shows FF data from \mbox{p-\=p} collisions at FNAL 
~\cite{cdf1}. The dotted line represents the lower limit for e-e FFs. There is a significant systematic difference between \mbox{p-p} and e-e FFs. The CDF FFs also reveal  suppression at larger parton energies relative to LEP e-e systematics.
Fig.~\ref{ttfig1} (fourth panel) is a surface plot of the p-p FF parametrization~\cite{evolve}---the e-e FF parametrization modified by cutoff factor
\bea
g_{cut}(y) = \tanh\{ (y - y_0)/\xi_y\}~~~y > y_0,
\eea
with $y_0 \sim \xi_y \sim 1.5$ determined by the CDF FF data~\cite{cdf1}. The cutoff  represents real fragment and energy loss from p-p relative to e-e FFs. The difference suggests that FFs may not be universal.

\begin{figure*}[t]
\includegraphics[height=1.65in]{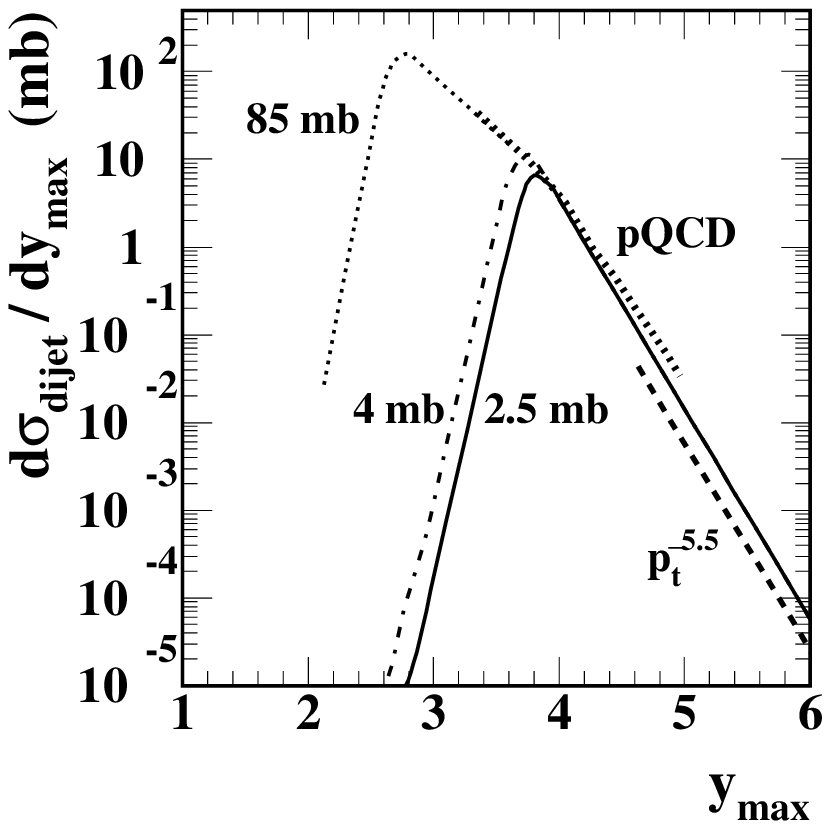} 
\includegraphics[height=1.67in]{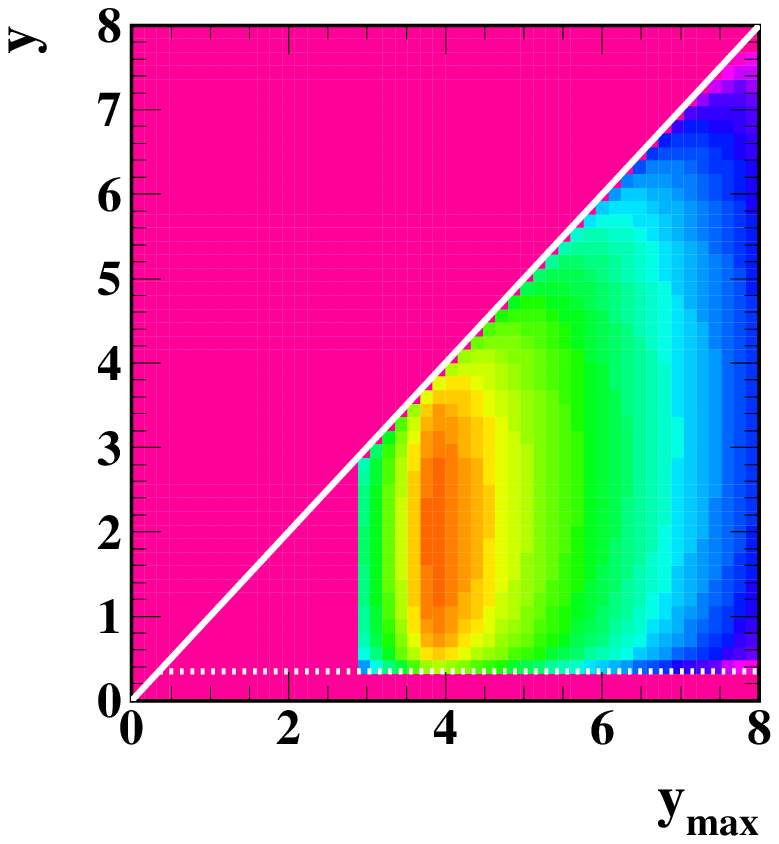}
\includegraphics[height=1.65in]{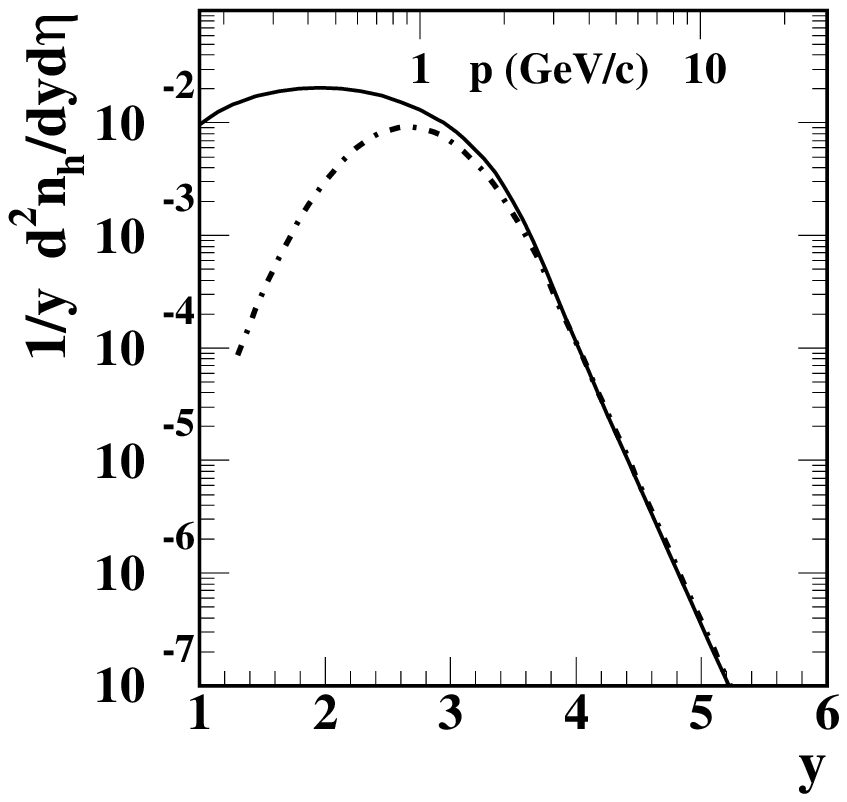}
\includegraphics[height=1.65in]{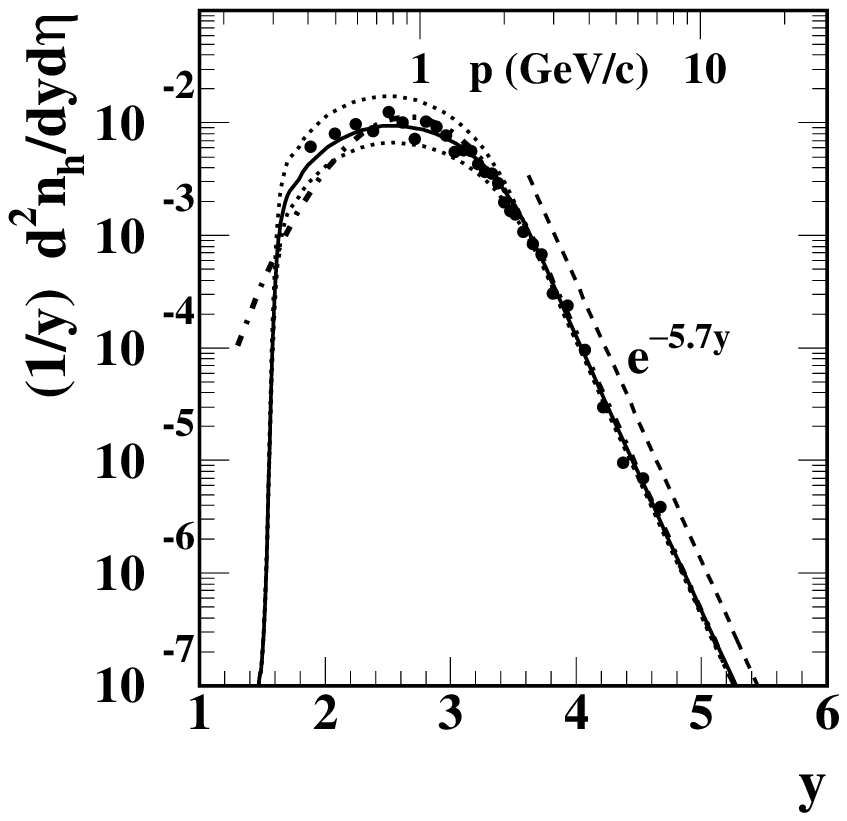}
\caption{\label{ttfig2}
First: Parton spectra inferred from this analysis for p-p collisions (solid curve) and central Au-Au collisions (dash-dotted curve) compared to an ab-initio pQCD theory result (bold dotted curve~\cite{cooper}),
Second: pQCD folding-integral argument for $e^+$-$e^-$ FFs,
Third: e-e FD (solid curve) and p-p hard-component reference (dash-dotted curve),
Fourth: Fragment distribution (solid curve) compared to p-p hard-component data (points). Dotted curves correspond to $\pm$10\% change in parton spectrum cutoff energy about 3 GeV. 
}
\end{figure*}

\section{$\bf p$QCD fragment distributions}

The parton $p_t$ spectrum from minimum-bias scattering into an $\eta$ acceptance near projectile mid-rapidity can be parametrized as
\bea
\frac{1}{p_t}\frac{d\sigma_{dijet}}{dp_t} &=& f_{cut}(p_t) \frac{A_{p_t}}{p_t^{n_{QCD}}} \rightarrow  \frac{d\sigma_{dijet}}{dy_{max}} \\ \nonumber
&& \hspace{-.7in} = f_{cut}(y_{max})\, A_{y_{max}}\, \exp\{-(n_{QCD} - 2)\, y_{max}\},
\eea
 with  $y_{max} \equiv\ln(2\,p_t / m_\pi )$. The cutoff factor
\bea
 f_{cut}(y_{max}) = \{ \tanh[(y_{max} - y_{cut})/\xi_{cut}] + 1\}/2
\eea
represents the minimum parton momentum which can lead to detectable charged hadrons as neutral pairs. Parton spectrum and cutoff parameters are determined by comparing FDs  with p-p and Au-Au spectrum hard components.

Fig.~\ref{ttfig2} (first panel) shows the parton spectrum (solid curve) with cutoff $\sim 3$ GeV inferred from a p-p $p_t$ spectrum hard component~\cite{evolve}. The bold dotted curve is an ab-initio pQCD calculation~\cite{cooper}. 
The spectrum integrates to $2.5 \pm 0.6$ mb, consistent with pQCD theory~\cite{sarc}. 

The pQCD folding (convolution) integral used to obtain fragment distributions is
\bea \label{fold}
\frac{d^2n_{h}}{dy\, d\eta}  \hspace{-.05in} &\approx&  \hspace{-.05in}  \frac{\epsilon(\delta \eta,\Delta \eta)}{ \sigma_{_{\tiny NSD}}\, \Delta \eta} \hspace{-.05in} \int_0^\infty \hspace{-.2in}  dy_{max} D(y,y_{max}) \frac{d\sigma_{dijet}}{dy_{max}},
\eea
where $D(y,y_{max})$ is the FF ensemble  from some collision system (e-e, p-p, A-A, in-medium or in-vacuum), and $d\sigma_{dijet}/dy_{max}$ is the parton spectrum~\cite{evolve}. Hadron spectrum hard component ${d^2n_{h}}/{dy\, d\eta}$ represents the fragment yield from  scattered parton pairs into  $\eta$ acceptance $\delta \eta$. Efficiency factor $\epsilon \sim 0.5$ 
includes the probability that the second jet also falls within $\delta \eta$. 
$\Delta \eta \sim 5$ is the effective $4\pi$ $\eta$ interval for scattered partons. $\sigma_{NSD}$ 
($\sim 36$ mb for $\sqrt{s_{NN}} = 200$ GeV) 
is the cross section for NSD p-p collisions.

Fig.~\ref{ttfig2} (second panel) shows integrand $D_{ee}(y,y_{max})\, \frac{d\sigma_{dijet}}{dy_{max}}$ of Eq.~(\ref{fold}) with unmodified FFs from e-e collisions and lower bound at  $y_{min} \sim 0.35$ ($p_t \sim 0.05$ GeV/c) (dotted line).
Fig.~\ref{ttfig2} (third panel) shows  the corresponding FD (
solid curve), 
the ``correct'' FD describing inclusive hadrons from partons produced by {\em free} parton scattering from p-p collisions. 
The dash-dotted curve is the hard-component model inferred from p-p spectrum data~\cite{ppprd}. The FD from e-e FFs lies well above the measured p-p hard component for hadron $p < 2$ GeV/c ($y < 3.3$), and the mode is shifted down to $\sim 0.5$ GeV/c. The ``correct'' e-e FD strongly disagrees with the hard component of the p-p $p_t$ spectrum. Nevertheless, the e-e FD is the proper reference for nuclear collisions~\cite{evolve}.

\begin{figure*}[t]
\includegraphics[height=1.65in]{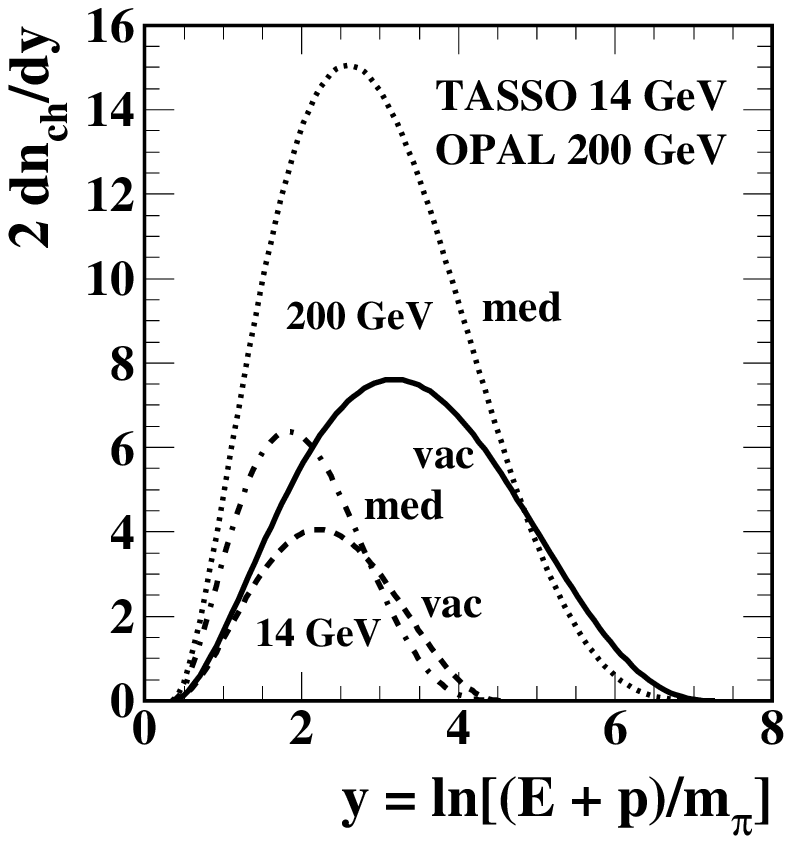} 
\includegraphics[height=1.67in]{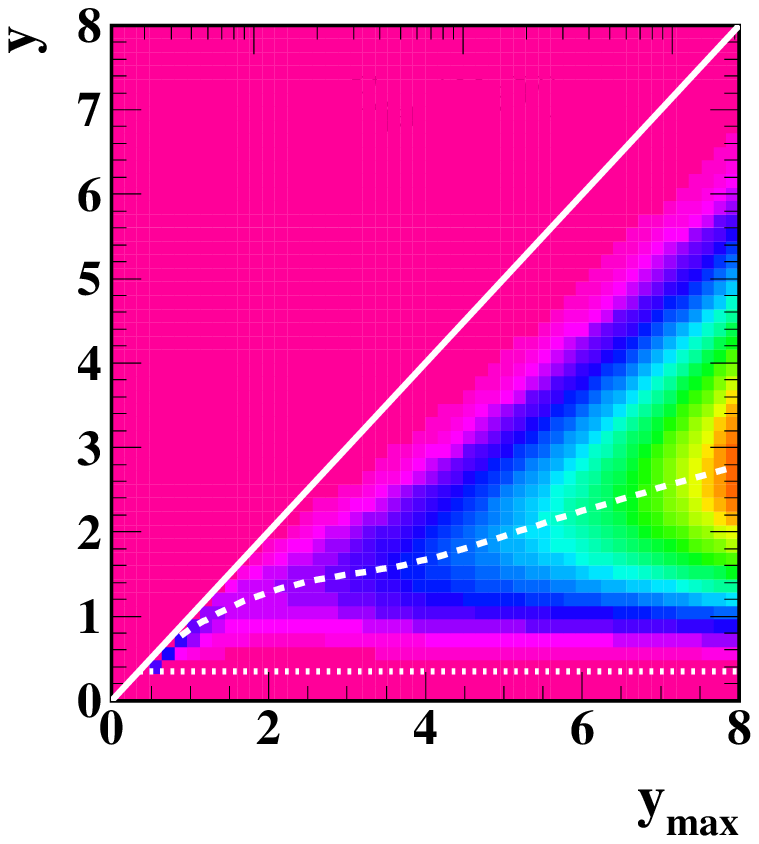}
\includegraphics[height=1.65in]{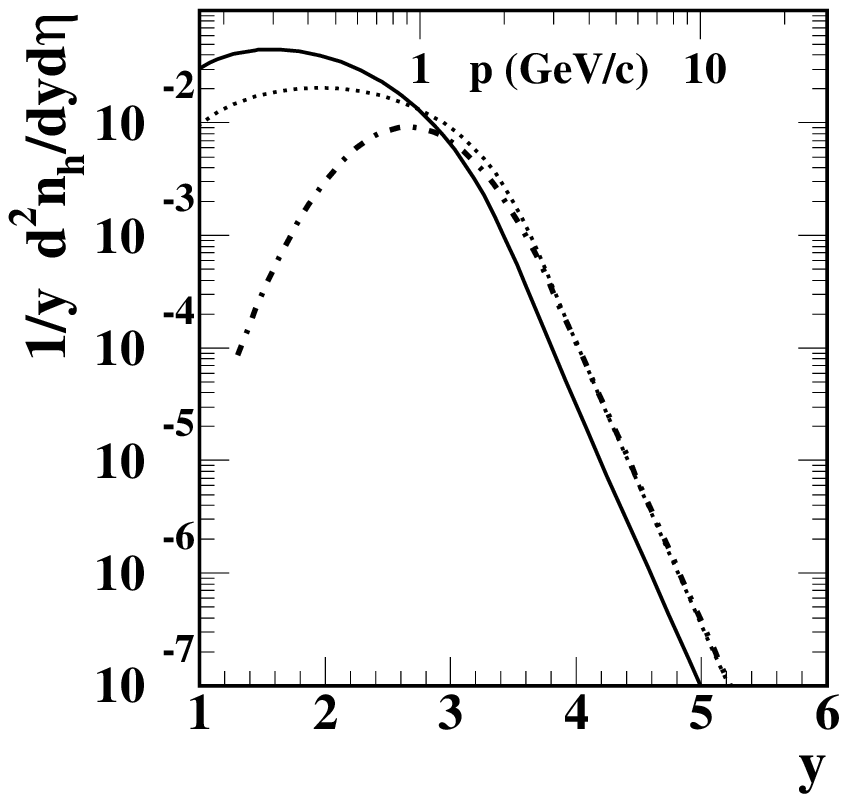}
\includegraphics[height=1.65in]{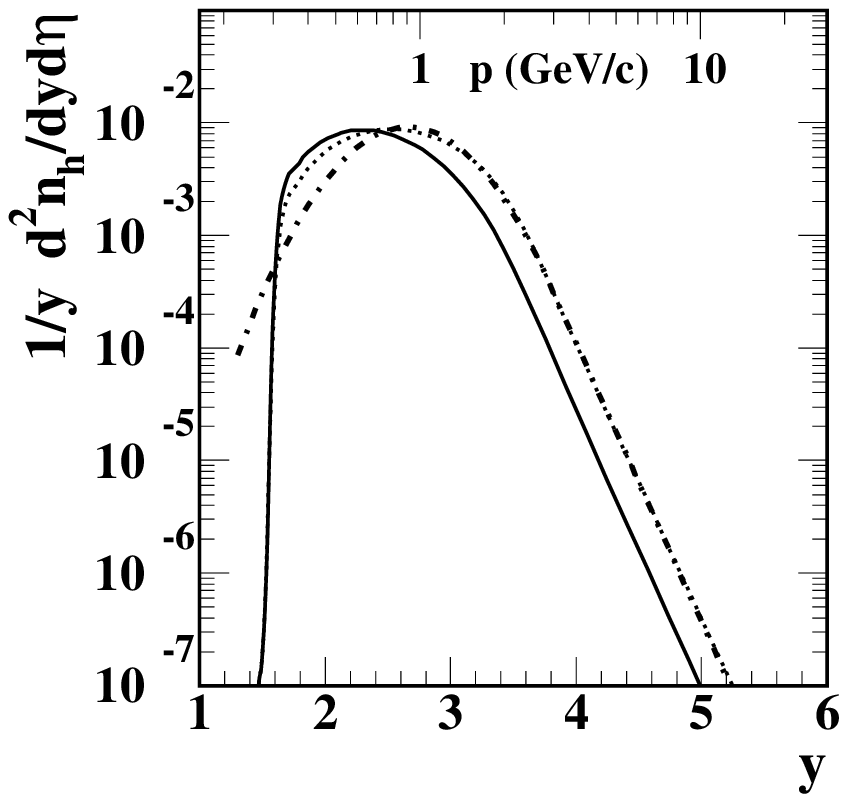}
\caption{\label{ttfig3}
First: $e^+$-$e^-$ FFs for two energies unmodified (solid and dashed curves) and modified to emulate parton ``energy loss''~\cite{bw} (dash-dotted and dotted curves),
Second: $e^+$-$e^-$ FF ensemble modified according to~\cite{bw},
Third: Medium-modified FD from $e^+$-$e^-$ FFs (solid curve) compared to in-vacuum $e^+$-$e^-$ FD (dotted curve)
Fourth: Medium-modified FD from p-\=p FFs (solid curve) compared to in-vacuum N-N FD (dotted curve).
}
\end{figure*}

Fig.~\ref{ttfig2} (fourth panel) shows FD $H_{NN-vac}$ as the solid curve, with measured FFs from \mbox{p-\=p} collisions. The mode of the FD is $\sim 1$ GeV/c. The solid points are hard-component data from p-p collisions and the dash-dotted curve is p-p model function $H_{pp}$~\cite{ppprd}. The comparison determines parton spectrum parameters $y_{cut} = 3.75$ ($E_{cut} \sim 3$ GeV), $A_{y_{max}}$ and exponent $n_{QCD} = 7.5$ and establishes a quantitative relationship among parton spectrum, measured FFs and measured spectrum hard components over all $p_t$, not just a restricted interval above 2 GeV/c.


\section{Parton ``energy loss'' model}      \label{eloss}


Fragmentation in A-A collisions requires a model of parton ``energy loss'' or medium modification. We adopt a minimal model of FF modification (Borghini-Wiedemann or BW)~\cite{bw}. 
%
 %
Figure~\ref{ttfig3} (first panel) illustrates the BW model (cf. Fig. 1 of~\cite{bw}, 
  $\xi_p = \ln(p_{jet}/ p)$ = $\ln(2\, p_{jet} / m_\pi) - \ln(2p/m_\pi) \sim y_{max} - y$).
In-vacuum e-e FFs for $Q = 14$ and 200 GeV from the beta parametrization are shown as dashed and solid curves~\cite{ffprd}. 
%
We can simulate BW accurately by changing parameter $q$ in $\beta(u;p,q)$ by $\Delta q \sim 1$
(dash-dotted and dotted curves)~\cite{evolve}. 
Small reductions at larger fragment momenta (smaller $\xi_p$) are compensated by much larger increases at smaller momenta. The largest changes (central Au-Au) correspond to an inferred  25\% leading-parton fractional ``energy loss.'' 
Fig.~\ref{ttfig3} (second panel) shows the modified e-e FF ensemble with FF modes shifted to smaller fragment rapidities $y$.

Figure~\ref{ttfig3} (third panel) shows $H_{ee-med}$ (solid curve), the FD obtained by inserting in-medium e-e FFs from the second panel into Eq.~(\ref{fold}). 
The dotted curve is the $H_{ee-vac}$ reference from in-vacuum e-e FFs. The mode of $H_{ee-med}$ is $\sim 0.3$ GeV/c.
Fig.~\ref{ttfig3} (fourth panel) shows results for \mbox{p-p} FFs. Major differences between p-p and \mbox{e-e} FDs appear below $p_t \sim 2$ GeV/c ($y_t \sim 3.3$). Conventional comparisons with theory (e.g., data {\em vs} NLO FDs) typically do not extend below 2 GeV/c~\cite{phenixnlo}. The large difference between the two collision systems below 2 GeV/c reveals that the small-$p_t$ region, conventionally assigned to hydro phenomena, may be essential for effective study of fragmentation evolution in A-A collisions.

\section{Fragmentation evolution}

\begin{figure*}[t]
\includegraphics[height=1.88in]{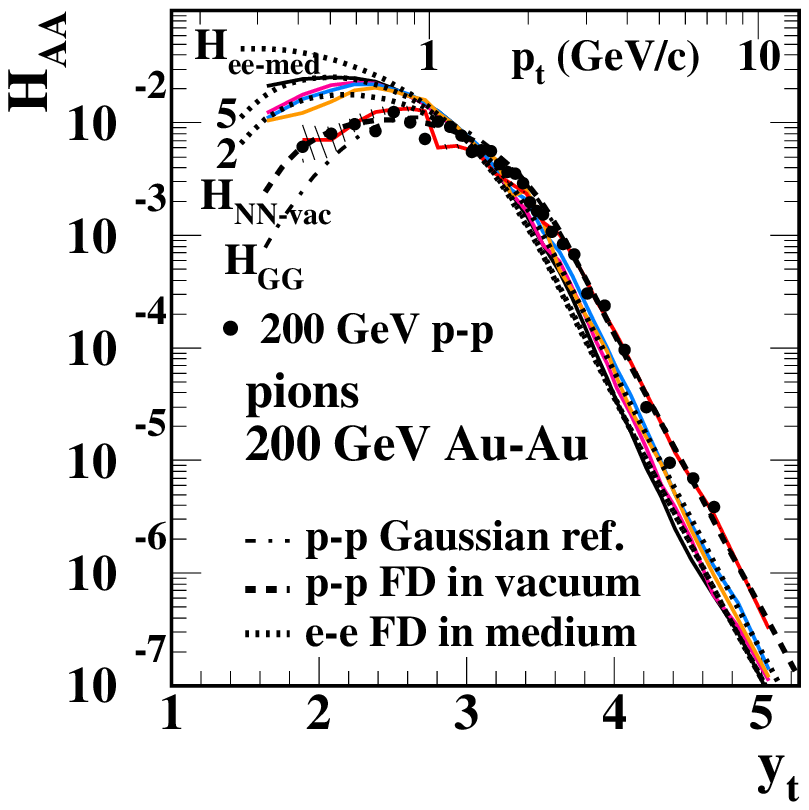}
\includegraphics[height=1.92in]{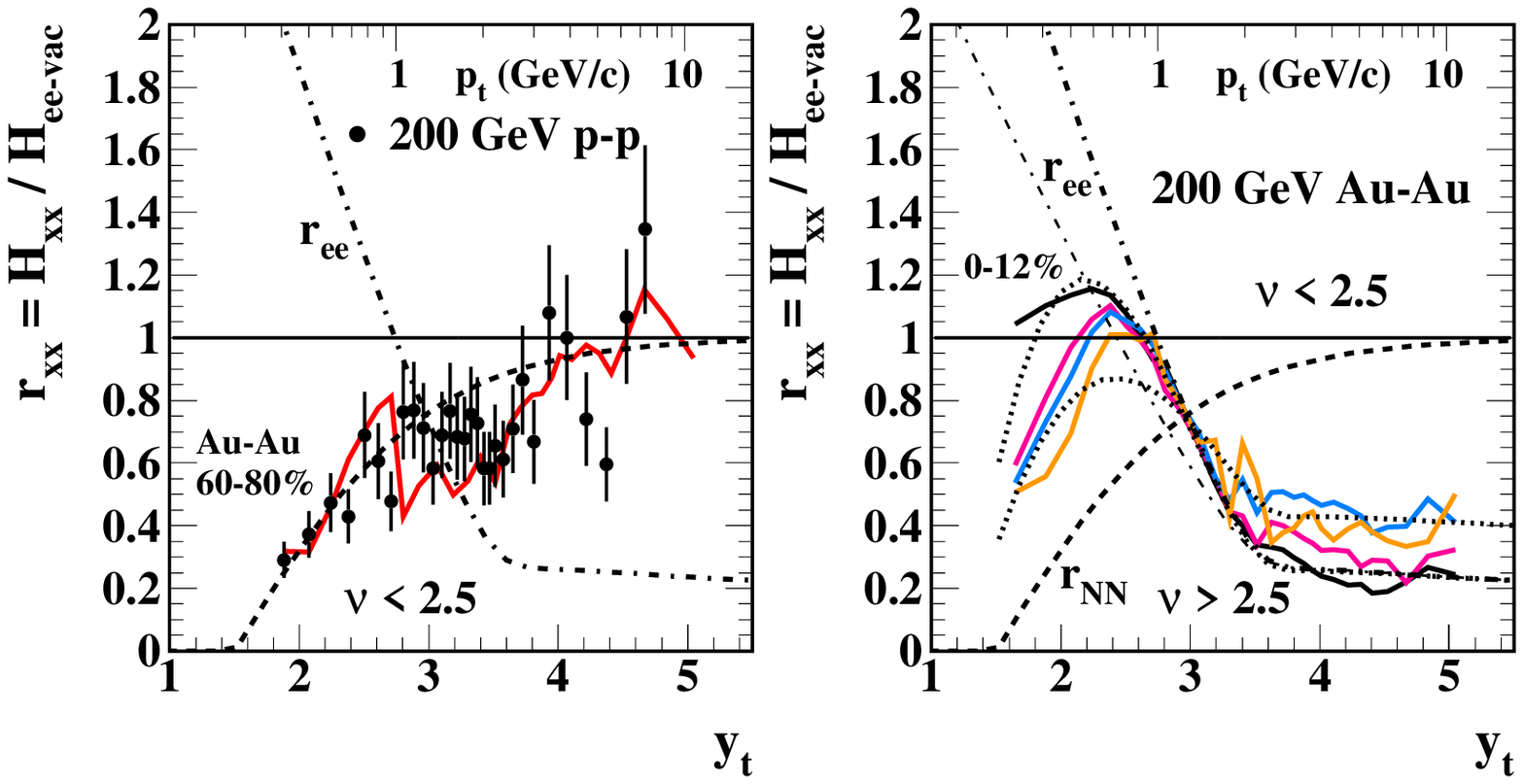} 
\caption{\label{ttfig4}
First: Hard-component evolution in central Au-Au collisions vs centrality~\cite{hardspec}. Large increases  at smaller $y_t$ accompany suppression at larger $y_t$.
Second: FD ratios relative to an ee-vacuum reference for Au-Au collisions below the sharp transition,
Third: FD ratios above the sharp transition revealing major changes in FD structure,
}
\end{figure*}

Measured FFs are combined with a parametrized pQCD parton spectrum to produce calculated $FD_{xx}$ for comparison with measured spectrum hard components $H_{xx}$. 
Figure~\ref{ttfig4} (first panel) shows spectrum hard components $H_{AA}$ (solid curves) for five centralities from 200 GeV Au-Au collisions~\cite{hardspec}. 
The hard components scale proportional to $n_{binary}$, as expected for parton scattering and fragmentation (jets). The points are from 200 GeV NSD p-p collisions~\cite{ppprd}. 
The dashed curve is $H_\text{NN-vac}$, and the upper dotted curve is $H_\text{ee-med}$ with $\Delta q = 1.15$, which corresponds to the most-central Au-Au curve (0-12\%). The parton spectrum cutoff for $H_\text{ee-med}$ has been reduced from 3 GeV ($y_{max} = 3.75$) to 2.7 GeV ($y_{max} = 3.65$) to match the central Au-Au hard component near $y_t = 3$. 

Jet-related spectrum structure can also be studied with ratios. The conventional spectrum ratio at RHIC is $R_{AA}$. Because  it includes the spectrum soft component $R_{AA}$ strongly suppresses fragment contributions at smaller $y_t$.
%
Hard-component evolution with centrality is better resolved by ratio $r_{AA} \equiv H_{AA} / H_{NN}$. 
%
However, studies in Ref.~\cite{evolve} reveal that
the proper  reference for all systems is the in-vacuum FD from e-e FFs, not p-p FFs. We therefore define ratios $r_{xx} = FD_{xx-yyy}/FD_{ee-vac}$ with xx = ee, NN, AA and yyy = med or vac to be compared with equivalent spectrum hard components $H_{xx-yyy}$.

Figure~\ref{ttfig4} (second panel) shows ratios redefined in terms of the ee-vac reference: $H_{pp}$ (\mbox{p-p} data -- points),  $H_{AA}$ (peripheral Au-Au data -- solid curve~\cite{hardspec}) and calculated $H_\text{ee-med}$ (dash-dotted curve) and $H_{NN-vac}$ (dashed curve) all divided by reference $H_{ee-vac}$. Strong suppression of p-p and peripheral Au-Au data apparent at smaller $y_t$ results from the cutoff of p-p FFs.

Figure~\ref{ttfig4} (third panel) shows measured $H_\text{AA}/H_\text{ee-vac}$ for more-central Au-Au collisions (solid curves) above a transition point on centrality at $\nu \sim 2.5$, 
with partial restoration of the suppressed region at smaller $y_t$ and strong suppression at larger $y_t$. The latter has been a major observation at RHIC 
(high-$p_t$ suppression, ``jet quenching''~\cite{starraa}). Newly apparent 
is the accompanying large {\em increase} in fragment yield {\em below} 2 GeV/c, still strongly correlated with the parent parton~\cite{daugherity}. 
Changes in fragmentation depend strongly on centrality near the transition point. It is remarkable that the trend at 10 GeV/c corresponds closely to the trend at 0.5 GeV/c. 
$H_{pp}$, $H_{AA}$ and ratios based on the e-e in-vacuum reference are  well described by pQCD FDs from 0.3 to 10 GeV/c~\cite{evolve}.

\section{Conclusions}

Hard components of $p_t$ spectra can be identified with minimum-bias parton fragmentation in nuclear collisions. Minimum-bias fragment distributions (FDs) can be calculated by folding a power-law parton energy spectrum with parametrized fragmentation functions (FFs) derived from $e^+$-$e^-$ and p-\=p collisions. Alterations to FFs due to parton ``energy loss'' or ``medium modification'' in Au-Au collisions are modeled by adjusting FF parametrizations consistent with rescaling QCD splitting functions. The reference for all nuclear collisions is the FD derived from in-vacuum $e^+$-$e^-$ FFs. Relative to that reference the hard component for p-p and peripheral Au-Au collisions is found to be {\em strongly suppressed} for smaller fragment momenta. At a specific point on centrality the Au-Au hard component transitions to enhancement at smaller momenta and suppression at larger momenta, consistent with FDs derived from  medium-modified $e^+$-$e^-$ FFs.

I thank the organizers of ISMD 2009 for a delightful and informative conference.
This work was supported in part by the Office of Science of the US DOE under grant DE-FG03-97ER41020.


\end{document}